\documentclass[a4paper,fleqn,usenatbib]{mnras}

\usepackage{mathptmx}

\usepackage[T1]{fontenc}
\usepackage{ae,aecompl}

\usepackage{graphicx}	
\usepackage{amsmath}	
\usepackage{amssymb}	

\usepackage{multirow}
\usepackage[flushleft]{threeparttable}

\title[Energy relaxation in collisions of C with O]{Energy relaxation in superthermal collisions of carbon with oxygen: the influence of isotopic substitution}

\author[Bop et al.]{
Cheikh T. Bop,$^{1}$\thanks{E-mail:cheikh-tidiane.bop@ucad.edu.sn}
M. Gacesa$^{1,2}$\thanks{E-mail: marko.gacesa@ku.ac.ae}
\\
$^{1}$
Department of Physics, Khalifa University of Science and Technology, PO Box 127788, Abu Dhabi, UAE \\
$^{2}$
Center for Catalysis and Separation, Khalifa University of Science and Technology, PO Box 127788, Abu Dhabi, UAE
}

\date{Accepted XXX. Received YYY; in original form ZZZ}

\pubyear{2025}

\begin{document}
\label{firstpage}
\pagerange{\pageref{firstpage}--\pageref{lastpage}}
\maketitle

\begin{abstract}
The transition from a once-dense Martian atmosphere to the thin one observed today implies a substantial loss of carbon, either through atmospheric escape or surface deposition. Accurately modeling this carbon escape necessitates accounting for collisions between energetic carbon atoms and the primary atmospheric constituents, including oxygen. To this end, we computed a highly accurate and comprehensive set of potential energy curves (PECs) for the C($^3$P) + O($^3$P) system. 
Based on these PECs, we derived statistically averaged total elastic and differential cross sections. 
Comparison with literature data for O($^3$P) + O($^3$P) collisions reveals that cross sections involving carbon can differ by up to a factor of two, indicating that oxygen is not a good proxy for modeling carbon escape. Furthermore, we evaluated the impact of all possible isotopic combinations in C($^3$P) + O($^3$P) collisions and found variations in cross sections of up to 8\%. Given the observed isotopic enrichment of carbon and oxygen in the Martian atmosphere, even such moderate differences can have a significant effect on escape models and the interpretation of planetary evolution.
\end{abstract}




%
\section{Introduction}
The evolution of the Martian atmosphere, from a potentially thick, water-sustaining early state to its current thin and arid conditions, is a fundamental problem in planetary science. This transition is largely attributed to the coupled effects of atmospheric escape to space and the sequestration of atmospheric carbon into the Martian surface \citep{kahn1985evolution,haberle1994model,jakosky2001mars,hu2015tracing}. While earlier models may have underestimated its significance, recent isotopic and geological evidence, including the identification of sedimentary carbonates in Gale Crater \citep{tutolo2025carbonates}, highlights a substantial carbonate sink that suggests Mars experienced a dynamic, self-regulating desert climate rather than sustained surface water, punctuated by transient ``oases'' \citep{Kite2025}. 

Parallel to surface processes, nonthermal escape mechanisms (\textit{i.e.}, photochemical escape, sputtering, and ion pickup) have been dominant in shaping Mars's atmospheric loss, driven by solar wind interactions and the planet's magnetic history \citep{jakosky2001mars}. Photochemical escape, in particular, involves energetic atomic species, such as carbon and oxygen fragments from CO$_2$ photodissociation \citep{2014Sci...346...61L}, capable of escaping Mars's gravity \citep{fox1999velocity,lammer2013outgassing,groller2014hot,lo2021carbon}. 

Quantifying the cumulative impact of these escape processes on atmospheric loss and isotopic evolution over geological timescales remains challenging, despite refined constraints from recent missions such as Mars Atmospheric Volatiles EvolutioN (MAVEN) \citep{jakosky2015mars,lillis2015escape,2018Icar..315..146J} and the Emirates Mars Mission \citep{amiri2022emirates,susarla2024variability}. MAVEN observations have directly measured upper-atmosphere atomic carbon densities, indicating that electron impact dissociation of CO$_2$ is a primary source of hot carbon atoms \citep{lo2022maven,sakai2024c+}. These findings suggest existing photochemical models may significantly underestimate carbon production and escape rates, underscoring the critical need for accurate modeling of collisional energy relaxation to model escape probabilities and associated isotopic fractionation.
The observed enrichment of heavier isotopes, such as $^{13}$C and $^{18}$O, in the Martian atmosphere \citep{krasnopolsky1996oxygen,2007Icar..192..396K,webster2013isotope,house2022depleted,alday2023photochemical} underscores the importance of incorporating isotopic variations into collisional dynamics.

The escape probabilities of superthermal carbon and oxygen atoms strongly depend on elastic and inelastic collision cross sections with ambient species, primarily CO$_2$, CO, N$_2$, and atomic oxygen, which mediate the energy relaxation. Nevertheless, detailed scattering studies of atomic carbon with major martian atmospheric constituents are notably lacking. Consequently, previous studies have often relied on approximating carbon collision cross sections using oxygen data such as, \textit{e.g.}, O($^3P$)+O($^3P$) \citep{kharchenko2000energy} or O($^3P$)+CO$_2$ \citep{2020MNRAS.491.5650G}, as a proxy \citep{lee2014hot,lo2021carbon,2024AGUFMP13D.3089L}, or mass-scaling relations (\textit{e.g.}, \citet{2014ApJ...790...98L}). Such approaches potentially misrepresent atomic carbon's true energy relaxation behavior in the upper atmosphere, leading to significant uncertainties in its escape rate estimates, as demonstrated for oxygen escape \citep{2014Icar..228..375F,lee2020effects}. A notable exception are recently reported elastic and inelastic cross sections for superthermal C($^3P$)+CO$_2$ scattering \citep{2024MNRAS.528.2621G}.

This work directly addresses the aforementioned gap by systematically investigating elastic collisions between C($^3$P) and O($^3$P), including all relevant isotopic combinations ($^{12}$C, $^{13}$C, $^{16}$O, $^{18}$O), to enhance the accuracy of carbon escape modeling and a more precise quantification of carbon loss and isotopic fractionation observed in the Martian atmosphere. 

This article is organized as follows: Section \ref{sec:comp} details the computational methods, Section \ref{sec:results} presents our cross section calculations and an analysis of the isotopic effects, and Section \ref{sec:conc} provides concluding remarks.

\section{Computational details}
\label{sec:comp}
The collisional processes we are interested in can be summarized as follows:
\begin{eqnarray}
    ^n{\rm C}(^3{\rm P})\ +\ ^m{\rm O}(^3{\rm P})\ \to\  ^n{\rm C}(^3{\rm P})\ +\ ^m{\rm O}(^3{\rm P}),
\end{eqnarray}
where $n=\{12,13\}$ and $m=\{16,17,18\}$. They rely on highly accurate potential energy curves (PECs) corresponding to the electronic states resulting from the interaction of C($^3$P) and O($^3$P). In practice, 18 adiabatic PECs, corresponding to singlet, triplet, and quintet multiplicities for $1\Sigma^+$, $2\Sigma^+$, $1\Pi$, $2\Pi$, $1\Delta$, and $1\Sigma^-$, are needed.
\subsection{Adiabatic potential energy curves}
The adiabatic PECs for the singlet and triplet electronic states correlating with the C($^3$P) + O($^3$P) asymptotic limit have been extensively studied through both theoretical and spectroscopic methods (see \cite{khalil2024theoretical} and references therein). In contrast, the quintet-state PECs have received considerably less attention. To the best of our knowledge, only \cite{bauschlicher1993theoretical} investigated all six quintet states using a high-level theoretical approach. While the PECs from \cite{khalil2024theoretical} are available online, those from \cite{bauschlicher1993theoretical} are not. Moreover, both sets of PECs are limited to intermediate internuclear separations (less than 8.5~$a_0$), making them unsuitable for scattering calculations, which are particularly sensitive to the long-range behavior of the potential. To address this limitation, we computed PECs for all 18 electronic states arising from the C($^3$P) + O($^3$P) asymptote using a highly accurate {\it ab initio} level of theory.

\paragraph*{}
The PECs were computed using the multireference configuration interaction method with single and double excitations (MRCISD). The Davidson correction (Q) has been applied to account for the effect of high-order corrections \citep{davidson1977size}. This method (MRCISD+Q), has been used as implemented in the \texttt{MOLPRO} (version 2015.1) quantum chemistry package \citep{werner2015version, werner2012molpro}. In all calculations, we employed the augmented correlation-consistent quintuple-zeta (AV5Z) Gaussian basis set of \citet{dunning1989gaussian}. The reference wavefunctions for the MRCISD+Q method were obtained from state-averaged complete active space self-consistent field (SA-CASSCF) calculations. To isolate states with specific values of the quantum number $\Lambda$—the projection of the electronic orbital angular momentum onto the molecular axis—we used \texttt{MOLPRO}’s built-in projector. Residual size-consistency errors were minimized by referencing all computed energies to the asymptotic limit of C($^3$P) + O($^3$P). The {\it ab initio} PECs were evaluated on a grid of internuclear distances ranging from 1.0 to at least 17.0~$a_0$.  The long-range interaction energies were calculated through extrapolation using the following inverse-power law:
\begin{eqnarray}
    V(R) = \frac{C}{R^{\lambda}},
\end{eqnarray}
where $C$ and $\lambda$ are the parameters determined considering the last {\it ab initio} points. We applied this methodology to compute the ground-state potential energy curve (PEC) correlating with the C($^3$P) + P($^4$S$^{\rm o}$) dissociation limit and derived spectroscopic constants in excellent agreement with experimental measurements \citep{Bop2017rotational}.  The complete set of {\it ab initio} points for the PECs correlating with the C($^3$P) + O($^3$P) dissociation limit is publicly available on Zenodo \citep{bop_zenodo}.

\subsection{Cross sections}
The theoretical framework for calculating atom-atom elastic scattering cross sections is well-established. The total elastic cross sections $\sigma$ for scattering atoms with collision energy $E$ are calculated as follows:
\begin{eqnarray}
    \sigma(E) = \sum_{l=0}^{\infty} \frac{4\pi}{k^2}(2l+1)sin^2(\eta_l),
\end{eqnarray}
where $k$ is the wave number and $\eta_l$ the phase shift for the $l$'th partial wave. The formula for differential cross sections through the polar angle $\theta$ into
element $d\Omega$ of solid angle is given by \cite{1933tac..book.....M} as follows:
\begin{eqnarray}
    \frac{d\sigma(E,\theta)}{d\Omega} = \left|\frac{1}{k}\sum_{l=0}^{\infty}(2l+1)sin(\eta_l)e^{i\eta_l}P_l(cos\theta)\right|^2,
\label{eq:diff_XS}
\end{eqnarray}
where $P_l$ denotes Legendre polynomial functions.  Therefore, the physical quantities that we are interested in depend on the phase shift which can be obtained by solving the differential equation
\begin{eqnarray}
    \left[\frac{d^2}{dR^2} - \frac{l(l+1)}{R^2} + k^2 -\left(\frac{2\mu}{\hslash^2}\right)V(R) \right]u_l(R) = 0,
\label{eq:schrod_atom}
\end{eqnarray}
where $\mu$ is the reduced mass of the collision system, $\hslash$ the reduced Planck's constant, $V(R)$ the potential for a given electronic state, and $R^{-1}u_l(R)$ the radial wave function. In practice, $\eta_l$ is derived through the boundary conditions:
\begin{eqnarray}
    u_l(R)_{R\to 0} & = & 0 \nonumber\\
    u_l(R)_{R\to \infty} & \sim & sin\left(kR - \frac{l\pi}{2} + \eta_l  \right)
\end{eqnarray}

In this work, $\eta_l$ and $\sigma$ were calculated using the \texttt{MOLSCAT} computer code which was initially designed for state-to-state non-reactive collisional processes involving molecules \citep{hutson1994molscat}. For the scattering of a molecule in a $^1\Sigma^+$ electronic state with a structureless atom ($^1S$), the Schr{\"o}dinger equation can be written in the form of equation~(\ref{eq:schrod_atom}) by setting the rotational quantum number of the molecule to zero and using one PEC which has no angular dependency. In this manner, it is possible to calculate highly accurate phase shifts and total elastic cross sections for atom-atom collisions using the \texttt{MOLSCAT} code. Indeed, we have tested this approach on the O($^3$P) + H($^2$S) collisions. By implementing the PECs computed by \cite{dagdigian2016accurate} in both the \texttt{MOLSCAT} and \texttt{VPA}\footnote{A recently released open-source code for atom-atom collisions.} codes \citep{palov2021vpa}, we derived total elastic cross sections that show excellent agreement with those reported by \cite{zhang2009energy}.

\begin{table}
    \centering
    \caption{Statistical weights for the C($^3$P) + O($^3$P) electronic states.}
    \label{tab:weights}
    \begin{tabular}{ccccc}
        \hline
        & $\Sigma^+$& $\Pi$&  $\Sigma^-$ & $\Delta$ \\
        \hline
        \hline
            Singlet & 1& 2& 1& 2\\
            Triplet & 3& 6& 3& 6\\
            Quintet & 5& 10& 5& 10\\
        \hline
    \end{tabular}
    \label{tab:my_label}
\end{table}
For the scattering of C($^3$P) + O($^3$P), we implemented each of the PECs that correlate with first asymptote in the \texttt{MOLSCAT} computer code to derive the corresponding total elastic cross sections and phase shifts. Equation~(\ref{eq:schrod_atom}) was solved using the propagator of \cite{alexander1987stable}.  The integrations were performed using a step size less than $10^{-3}~a_0$ from 1.0 to 8~$a_0$ which was smoothly increased for larger distances to account for the asymptotic behavior of the PECs. The number of partial waves was determined by setting the convergence of the total elastic cross sections to $10^{-4}~\mathring{A}^2$. For each PEC, the collision energy ranges from 0.01 to 5.0~eV. The differential cross sections were computed separately by means of equation~(\ref{eq:diff_XS}). The calculations were performed in steps of $0.1^{\circ}$ for angles ranging from 0 to 180 degrees. 

The elastic cross sections (differential cross sections) obtained using the different PECs, were then averaged employing the statistical weights as shown in Table~\ref{tab:weights}. To account for isotopic substitution, we used reduced masses of 6.856, 7.172, 7.034, 7.368, 7.200, and 7.549~au for $^{12}$C$^{16}$O, $^{13}$C$^{16}$O, $^{12}$C$^{17}$O, $^{13}$C$^{17}$O, $^{12}$C$^{18}$O, and $^{13}$C$^{18}$O, respectively. These collision systems are hereafter denoted 26, 36, 27, 37, 28, and 38, where the label is constructed by taking the last digit of the total number of nucleons of C and O atoms, \textit{e.g.}, 26 for $^{12}C^{16}$O, 37 for $^{13}C^{17}$O, etc.

\section{Results and discussions}
\label{sec:results}

\begin{figure}
    \centering
    \includegraphics[width=0.98\linewidth, trim = 0 0 40 38, clip = true]{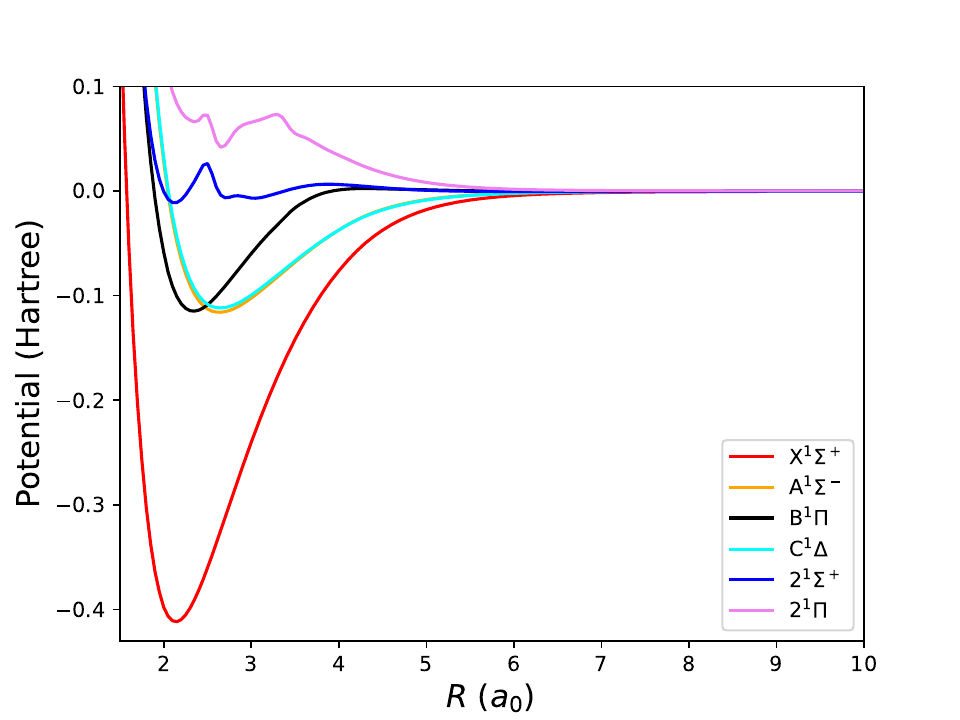}
    \includegraphics[width=0.98\linewidth, trim = 0 0 40 38, clip = true]{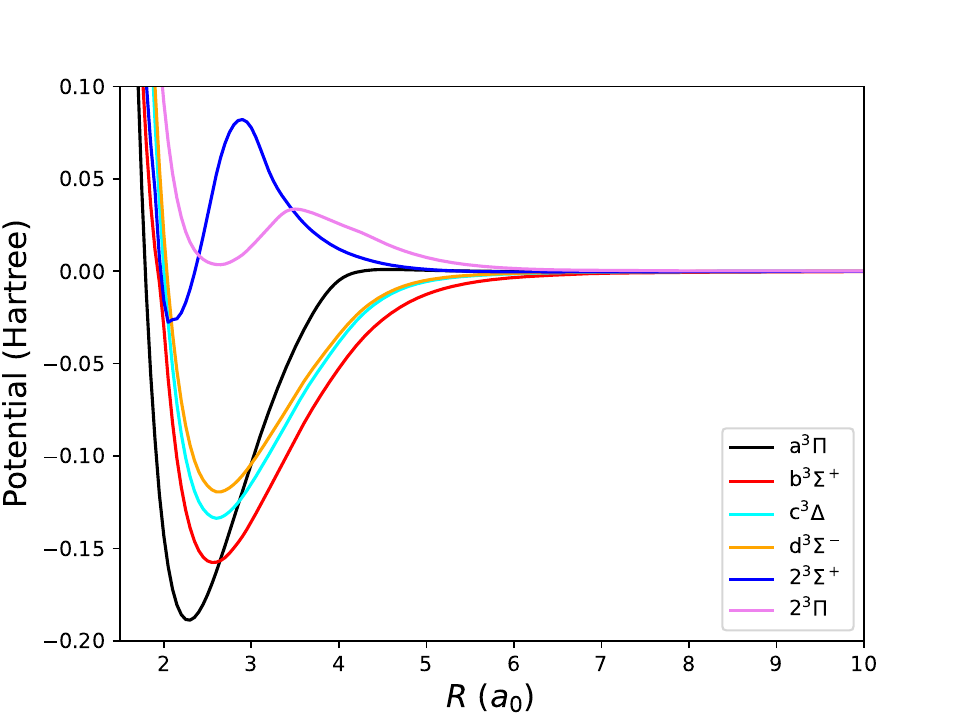}
    \includegraphics[width=0.98\linewidth, trim = 0 0 40 38, clip = true]{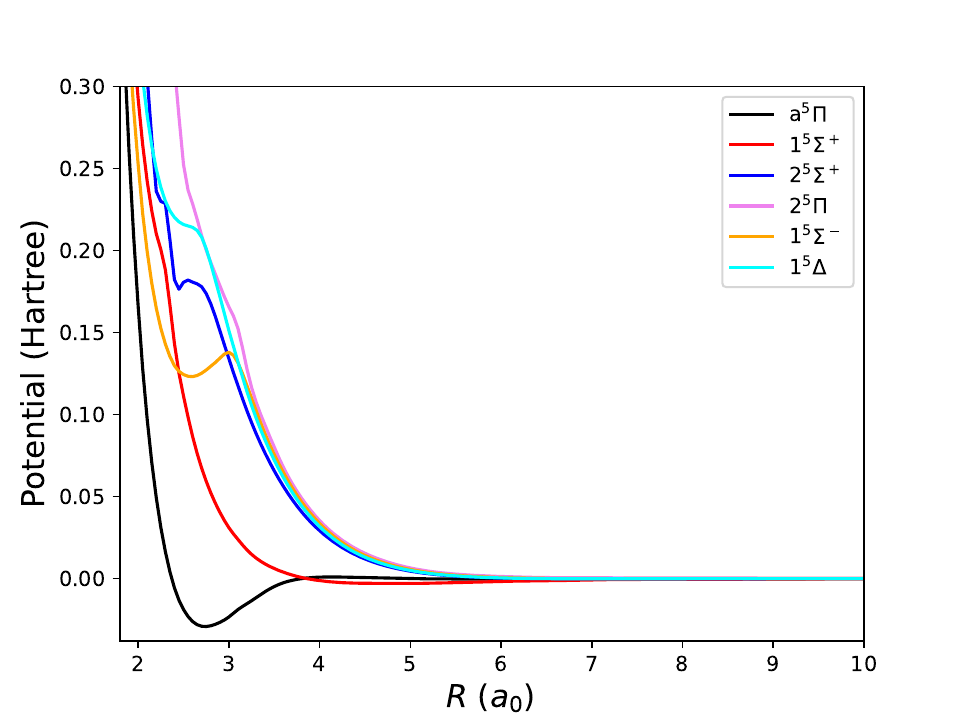}
    \caption{Potential energy curves for CO electronic states that correlate with the first asymptote.}
    \label{fig:pec}
\end{figure}

We display in Fig.~\ref{fig:pec} the PECs for the 18 electronic states correlating with C($^3$P) + O($^3$P). The behavior of the singlet and triplet states, including the double minimum of the 2$^1\Sigma^+$ state, has been extensively discussed in the literature (see \cite{khalil2024theoretical} and references therein). The detailed shape of the PECs is beyond the scope of this work and will not be discussed further in this manuscript.

In terms of magnitude, our PECs are consistently lower in energy than reported by \citet{khalil2024theoretical}. The qualitative agreement, despite the energy shift, can be attributed to the lower-quality triple-zeta basis set used in the work of \citet{khalil2024theoretical} compared to the quintuple-zeta basis set used in our calculations.Furthermore, our results suggest that these authors may have misidentified the b$^3\Sigma^+$ state as a $^3\Pi$ state and reported three states (labeled a$^3\Pi$, a'$^3\Pi$, and k$^3\Pi$) instead of the expected two $^3\Pi$ states correlating with the C($^3$P) + O($^3$P) asymptote.

For the quintet states, our results reproduce all the qualitative features of the PECs reported by \cite{bauschlicher1993theoretical}, but again are shifted consistently lower in energy. For example, the inner potential well of the a$^5\Pi$ state in our calculation is approximately 386~cm$^{-1}$ deeper. This discrepancy is likely due to the relatively small basis sets, triple zeta for the $\Sigma^-$ and $\Delta$ states and quadruple zeta for the $\Sigma^+$ and $\Pi$ states, used in that study.

\begin{figure*}
    \centering
    \includegraphics[width=0.32\linewidth, trim = 5 0 45 35, clip = true]{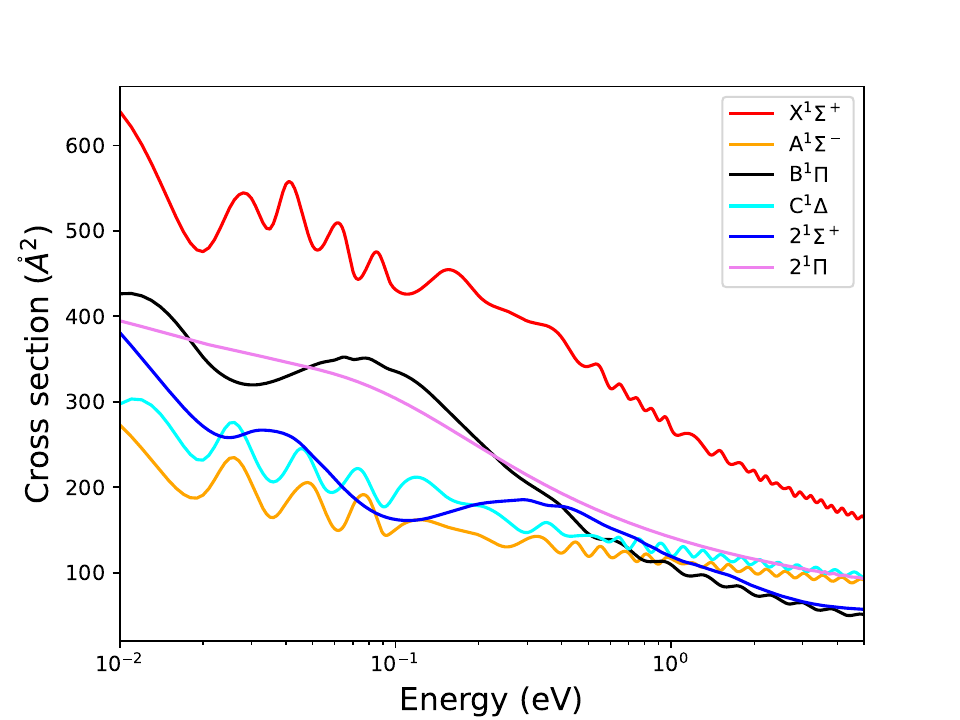}
    \includegraphics[width=0.32\linewidth, trim = 5 0 45 35, clip = true]{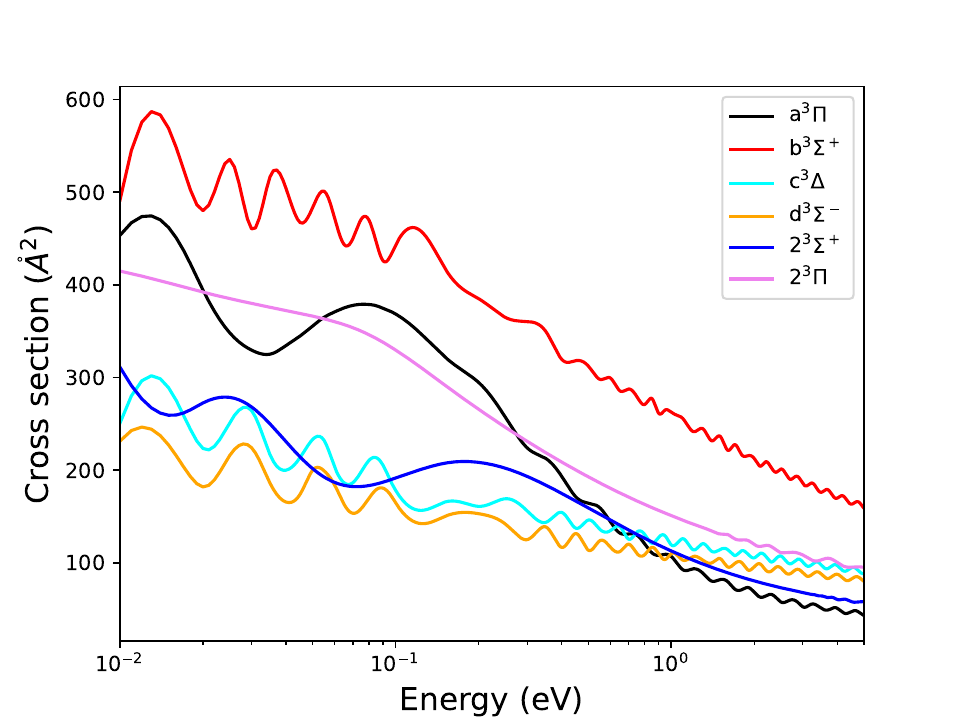}
    \includegraphics[width=0.32\linewidth, trim = 5 0 45 35, clip = true]{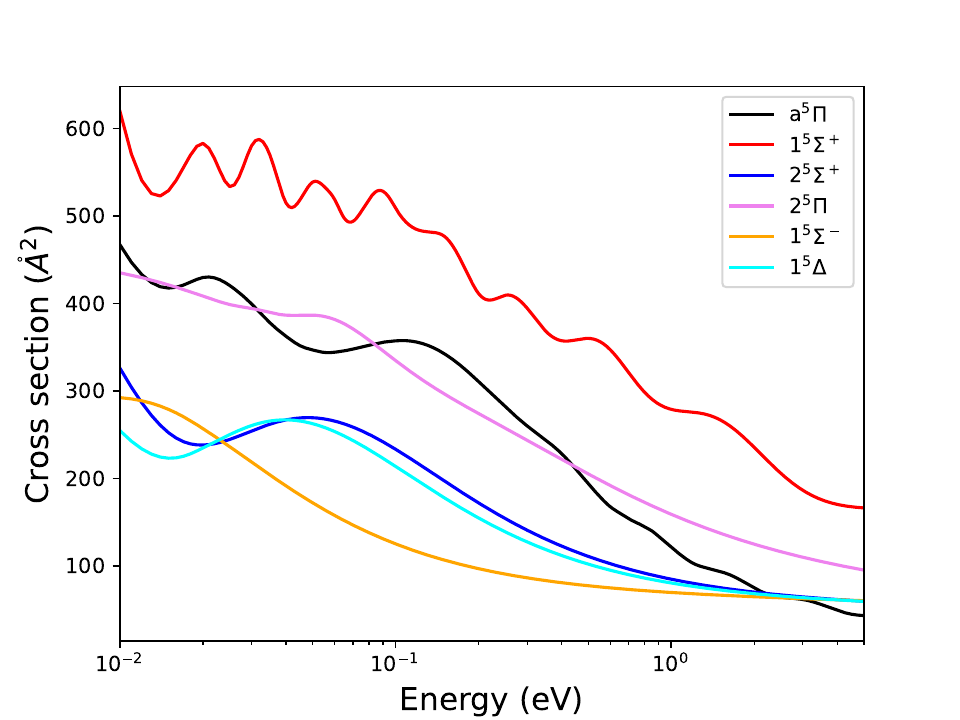}
    \caption{Total elastic cross sections for the different CO electronic states, correlating with the first
asymptote, as a function of kinetic energy. The labels of the cross sections are derived from the electronic states.}
    \label{fig:txs}
\end{figure*}

In summary, we have calculated the most complete and accurate set of PECs for the electronic states correlating with the C($^3$P) + O($^3$P) dissociation limit. These PECs, computed using a quintuple-zeta basis set, encompass all 18 states and provide an accurate description of the long-range potential, which is essential for scattering calculations.

Figure~\ref{fig:txs} shows the total elastic cross sections computed using the PECs of the 18 electronic states correlating with the C($^3$P) + O($^3$P) dissociation limit. The observed oscillations in the cross sections are characteristic of glory scattering, arising from molecular potentials with both attractive and repulsive regions. In contrast, smoother profiles are found for channels dominated by repulsive interactions, such as the $2^{1,3,5}\Pi$ and $1^5\Sigma^-$ states \citep{zygelman1994molecular}.

\begin{figure}
    \centering
    \includegraphics[width=0.98\linewidth, trim = 5 0 42 35, clip = true]{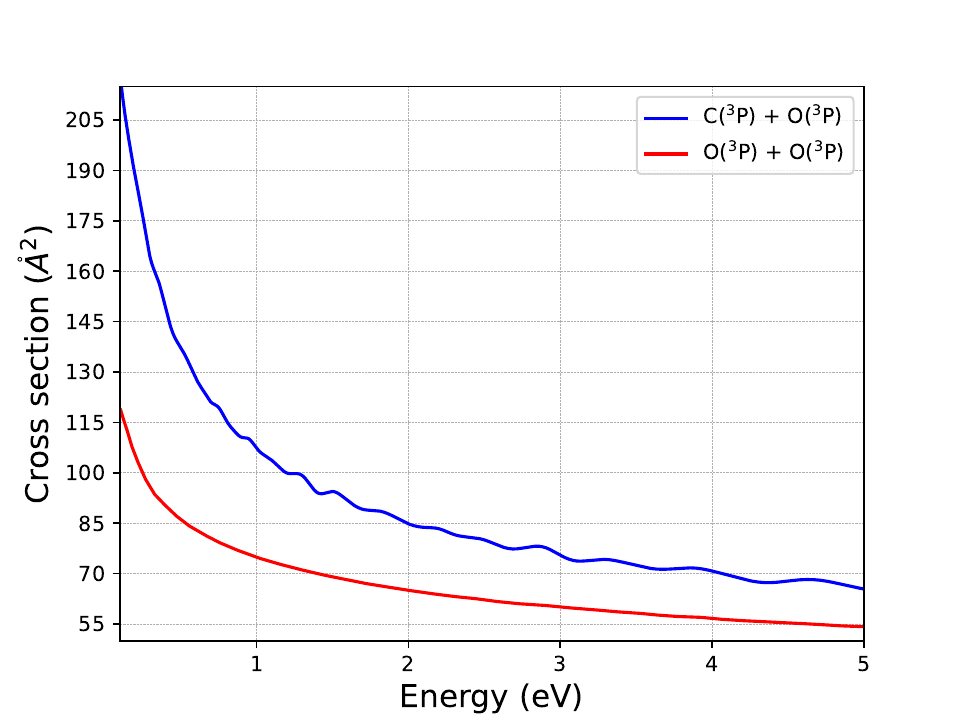}
    \caption{Dependence on kinetic energy of the statistically averaged cross sections induced by C($^3$P) + O($^3$P) collisions (this work) and O($^3$P) + O($^3$P) scattering \citep{kharchenko2000energy}.}
    \label{fig:CO_tot_XS}
\end{figure}

The statistically averaged total elastic cross sections induced by C($^3$P) + O($^3$P) collisions are compared in Fig.~\ref{fig:CO_tot_XS} to those due to O($^3$P) + O($^3$P) scattering computed by \cite{kharchenko2000energy}. At low energy,  the C($^3$P) + O($^3$P) total elastic cross sections are larger than those of O($^3$P) + O($^3$P) by nearly a factor of two. The difference decreases as the energy increases and reaches a value of $\sim10~\mathring{A}^2$ at 5~eV. 
The substantial differences, namely a factor of two at low energy and non-linear behavior of the cross sections, confirm that using O($^3$P) + O($^3$P) as a simple proxy for C($^3$P) + O($^3$P) collisions is not a valid approach for accurately modeling carbon escape. 

\begin{figure}
    \centering
    \includegraphics[width=0.98\linewidth, trim = 20 10 10 10, clip = true]{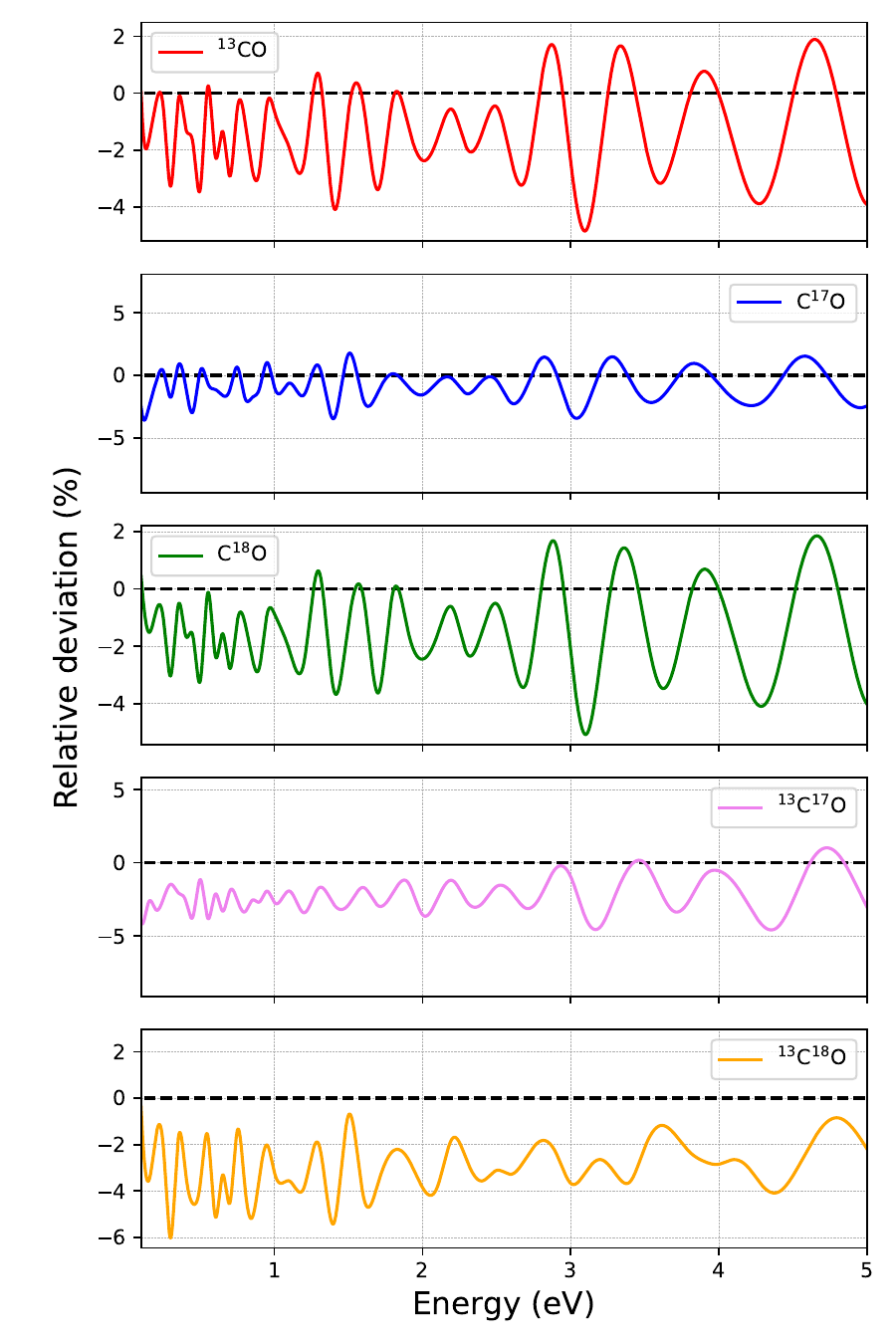}
    \caption{Kinetic energy dependence of the relative deviations upon isotopic substitution compared to CO statistically averaged total elastic cross sections.}
    \label{fig:iso_total_XS}
\end{figure}

We display in Fig.~\ref{fig:iso_total_XS} the effect of isotopic substitution on the statistical average total elastic cross sections induced by C($^3$P) + O($^3$P) collisions. As one can see, the relative deviations are globally shifted downward at high energies, mainly reflecting underestimations of the reference cross sections. 

The energy-dependent oscillations in the relative deviations, which vary by up to 8\%, demonstrate that the isotopic effect in C($^3$P) + O($^3$P) collisions cannot be captured by simple mass scaling relations. This underscores the necessity of dedicated, high-accuracy calculations for each isotopologue to correctly model isotopic fractionation in the Martian atmosphere.
The amplitude of the oscillations varies from -5\% to 2\%, -8\% to 7\%, -6\% to 2\%, -8\% to 5\%, and -6\% to 2\% for the ``36'', ``27'', ``28'', ``37'', and ``38'' collision systems, respectively.
Therefore, the influence of isotopic substitution in the C($^3$P) + O($^3$P) scattering varies up to 8\% depending on the kinetic energy and the isotopes. Similar effects have been reported for isotopic substitution in Ar + O$_2$ collisions \citep{Bop2021inelastic}, as well as in superthermal atom-diatom collisions involving atomic O($^3$P) \citep{chhabra2023quantum,2014JChPh.141p4324G}. 

\begin{figure}
    \centering
    \includegraphics[width=0.98\linewidth, trim = 8 2 35 35, clip = true]{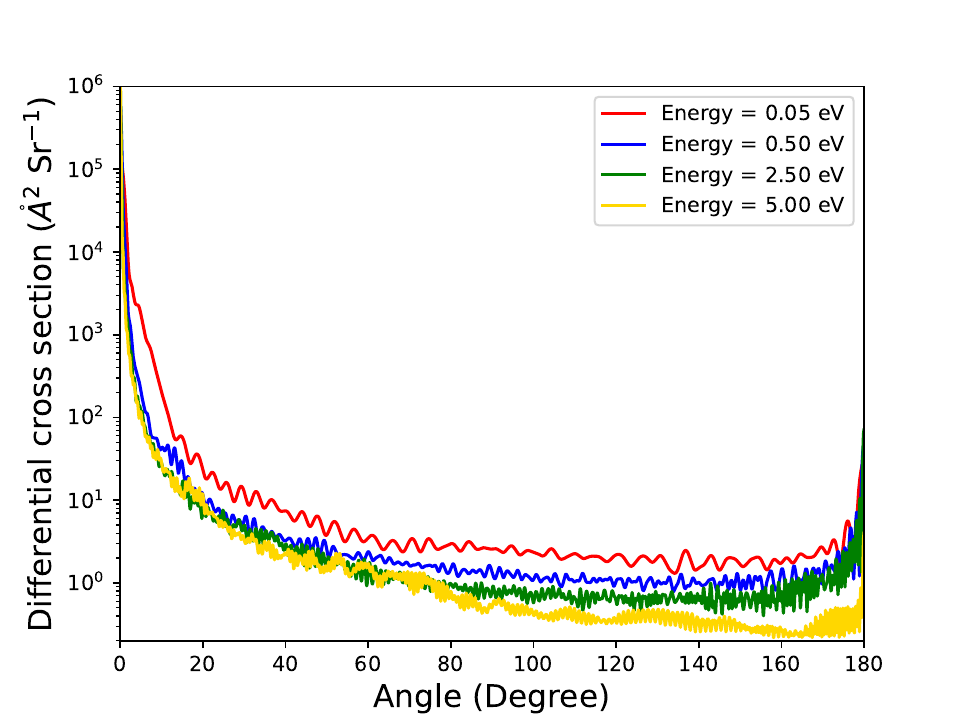}
    \caption{Angular dependence of statistically averaged differential cross sections, for kinetic energies of 0.05, 0.5, and 5.0~eV, for the CO collision system.}
    \label{fig:CO_dif_XS}
\end{figure}

Fig.~\ref{fig:CO_dif_XS} presents the statistically averaged differential cross sections for C($^3$P) + O($^3$P) collisions at selected energies. All curves exhibit forward peaking, a well-known physical behavior observed in differential cross sections, primarily due to the contribution of long-range interaction potentials in the collisional process. Despite the numerous oscillations -- likely present because of a denser angular mesh used in this work -- the overall behavior of these cross sections is similar to that reported for O($^3$P) + O($^3$P) collisions by \cite{kharchenko2000energy}. For isotopic substitutions, the differential cross sections exhibit a comparable shape and are therefore not shown here. The relative deviations from the main collision system are of the same order of magnitude as those shown in Fig.~\ref{fig:iso_total_XS}.

\section{Conclusion}
\label{sec:conc}
We calculated a complete and highly accurate set of potential energy curves (PECs), including all spin symmetries and a proper description of the long-range interaction, correlating with the C($^3$P) + O($^3$P) dissociation limit. We used the PECs to derive statistically averaged total elastic and differential cross sections. The resulting total elastic cross sections for C($^3$P) + O($^3$P) collisions are greater than those for O($^3$P) + O($^3$P) scattering \cite{kharchenko2000energy} by up to a factor of two. Therefore, using oxygen-atom collision data to model carbon escape in the Martian atmosphere may introduce a significant source of error in estimates of the total escape rate of carbon possibly impacting the early martian climate scenarios \citep{2019LPICo2089.6033L}. The statistically averaged differential cross sections reported in this work may be directly applied to evaluate transport coefficients such as diffusion, viscosity, and thermal conductivity in C + O mixtures relevant to modeling of carbon transport and escape in CO$_2$-rich planetary and exoplanetary atmospheres. The same collisional dynamics are relevant to understanding the photochemistry of Earth's upper atmosphere, where collisions between atomic oxygen and carbon are key processes that mediate energy transfer, drive chemical reactions, and contribute to phenomena like airglow \citep{2000GL012189}.

Considering all possible isotopic combinations in the C($^3$P) + O($^3$P) collisions, we estimated the resulting effect on the cross sections to be as high as 8\%. In light of the observed enrichment of carbon and oxygen isotopes in the Martian atmosphere, even this relatively small variation may significantly impact models of carbon escape and carbon isotope fractionation at Mars \citep{lo2021carbon,yoshida2023strong,2023PSJ.....4...41T}. We recommend using the data reported in this work for accurate modeling of carbon escape and related collisional processes in the Martian and Venusian atmospheres, where isotopic effects are important for a better understanding of their temporal evolution.

\section*{Acknowledgments}
The authors acknowledge Khalifa University of Science and Technology for funding this work through project \#8474000740‐RIG‐2024‐045. The data will be shared on a request to the
corresponding authors.

\section*{Conflicts of interest}
There are no conflicts to declare.

\section*{Data availability}
The data underlying this article are available on Zenodo \citep{bop_zenodo}.


\bibliographystyle{mnras}
\bibliography{biblio}





\bsp	
\label{lastpage}
\end{document}